# Influence Factors of the Evaporation Rate of a Solar Steam Generation System: a Numerical Study

Jinxin Zhong[1], Congliang Huang[1,2a)], Dongxu Wu[1], and Zizhen Lin[1]

[1] School of Electrical and Power Engineering, China University of Mining and Technology, Xuzhou 221116, P. R. China.
[2] Department of Mechanical Engineering, University of Colorado, Boulder, Colorado 80309-0427, USA.
[a)] Author to whom correspondence should be addressed. E-mail: huangcl@cumt.edu.cn.

**Abstract:** Many efforts have been dedicated to improve the solar steam generation by using a bi-layer structure. In this paper, a two-dimensional mathematical model describing the water evaporation in a bi-layer structure is firstly established and then the finite element method is used to simulate the effects of different influence factors on the evaporation rate. Results turn out that: besides the high solar energy absorptivity of the first-layer, an optimum porosity of the second-layer porous material should be applied and the optimum porosity is about 0.45 in this work. This optimum porosity is determined by the balance between the positive effect of the lowering effective thermal conductivity of the second layer and the negative effect of the reduced vapor diffusivity in the second layer when the porosity is decreased. The influence of the thermal conductivity of the second-layer porous material is negligible because the effective thermal conductivity of the second layer is determined by the porosity while a larger porosity means more water in the second layer. The ambient air velocity could greatly enhance the evaporation rate, and the evaporation rate will decrease linearly with the increase of the air relative humidity. This study is expected to supply some information for developing a more effective bi-layer solar steam generation system.

**Keywords: solar steam generation; solar energy; numerical method; porous material**

*Nomenclature*

| | | | |
|---|---|---|---|
| $a_w$ | Water activity | $R$ | Ideal gas constant in Eq.(6), J·mol$^{-1}$·K$^{-1}$ |
| $c$ | Concentration, mol·m$^{-3}$ | $R$ | Reaction rate, mol·m$^{-3}$·s$^{-1}$ |
| $c_{sat}$ | Saturation vapor concentration, mol·m$^{-3}$ | $S$ | Saturation |
| $c_{v0}$ | Initial water vapor concentration, mol·m$^{-3}$ | $S_{iw}$ | Initial water saturation of second layer |
| $C_p$ | Heat capacity, J·kg$^{-1}$·K$^{-1}$ | $T$ | Temperature, K |
| $D_{cap}$ | Capillary diffusivity, m$^2$·s$^{-1}$ | $\boldsymbol{u}$ | Velocity, m·s$^{-1}$ |
| $D_{eff}$ | Effective vapor diffusivity, m$^2$·s$^{-1}$ | $\boldsymbol{u}_{mean}$ | Fluid velocity in second layer, m·s$^{-1}$ |
| $D_{va}$ | Air-vapor diffusivity, m$^2$·s$^{-1}$ | *Greek letters* | |
| $e_b$ | Blackbody hemispherical total emissive power, W·m$^{-2}$ | $\varphi_p$ | Porosity |
| $G$ | Irradiation, W·m$^{-2}$ | $\mu$ | Viscosity, kg·m$^{-1}$·s$^{-1}$ |
| $H_{vap}$ | Latent heat of evaporation, J·kg$^{-1}$ | $\varepsilon$ | Surface emissivity |
| $\boldsymbol{I}$ | Identity matrix | $\kappa$ | Permeability of the porous matrix, m$^2$ |
| $k$ | Thermal conductivity, W·m$^{-1}$·K$^{-1}$ | $\kappa_r$ | Relative permeability, m$^2$ |
| $k_e$ | Effective Thermal conductivity, W·m$^{-1}$·K$^{-1}$ | $(\rho C_p)_e$ | Effective volumetric heat capacity, J·m$^{-3}$·K$^{-1}$ |
| $K$ | Evaporation coefficient, l·s$^{-1}$ | $\rho$ | Density, kg·m$^{-3}$ |
| $L_{in}$ | Entrance length, m | *Subscripts* | |
| $M$ | Molecular weight, kg·mol$^{-1}$ | 0 | Ambient initial conditions |
| $\boldsymbol{n}$ | Normal vector | $a$ | Dry air |
| $\boldsymbol{n}$ | Mass flux in Eq.(20), kg·m$^{-2}$·s$^{-1}$ | $g$ | Gas in second layer. |
| $P$ | Pressure, Pa | $s$ | Solid container |
| $P_{in}$ | Entrance pressure, Pa | $wg$ | Vapor in second layer |
| $P_{out}$ | Exit pressure, Pa | $v$ | Air in ambient air domain |
| $P_{sat}$ | Saturation pressure in second layer, Pa | $w$ | Water in second layer |
| $\boldsymbol{q}$ | Heat flux vector, W·m$^{-2}$ | $tot$ | Fluid of second layer |
| $\boldsymbol{q}_e$ | Radiation heat flux, W·m$^{-2}$ | $p$ | Porous second layer |
| $Q_{vap}$ | The heat of evaporation, W·m$^{-3}$ | | |

1. Introduction

Solar-enabled evaporation is a solar energy harvesting technology which has a potential application in modern power plants, chemical plants, and seawater desalination plants [1-9]. Recently, by applying a bi-layer structure for solar steam generation which was firstly proposed by Ghasemi et al. [10], many efforts have been dedicated to minimize the heat losses and thus to improve the evaporation efficiency of water. In a bi-layer structure, the first layer is the absorbent material used to absorb sunlight, and the second layer is a porous material for water absorption and heat insulation. A large number of interests have been focused on finding novel first-layer nanomaterials for high efficiency of light absorbing, such as, gold nanoparticle [11-14], graphene [15, 16], carbon black nanoparticles [17, 18], carbon foam [10, 19], activated carbon[20] and carbon nanotubes [21]. The porous material is usually applied as the second layer material because of its high capillary effect and low thermal conductivity [22-26], such as wood [27, 28], carbonized wood [29], carbonized mushrooms [30], and cellulose nanofibers [31]. Although a large number of studies have been carried out to look for cheaper and more efficient materials to increase the evaporation rate of the bi-layered structure for solar steam generation, to our knowledge, there is still no detailed study about the influence factors of the evaporation process in a bi-layer system, especially the ambient air velocity, the porosity of the second layer, etc.

In this paper, a two-dimensional mathematical model describing the water evaporation in a bi-layer system is firstly established, and then the finite element method is used to simulate the effects of different influence factors on the evaporation process, such as the ambient air velocity, the air relative humidity, the absorptivity of the first-layer material, the porosity and the thermal conductivity of the second-layer material. It shows that the thermal conductivity of the second layer is not as important as it assumed to be [32], and there is an optimum porosity for the second layer to enhance the evaporation rate. This study is expected to supply some information for developing a more effective bi-layer solar steam generation system.

2. Mathematical model

2.1 Physical model

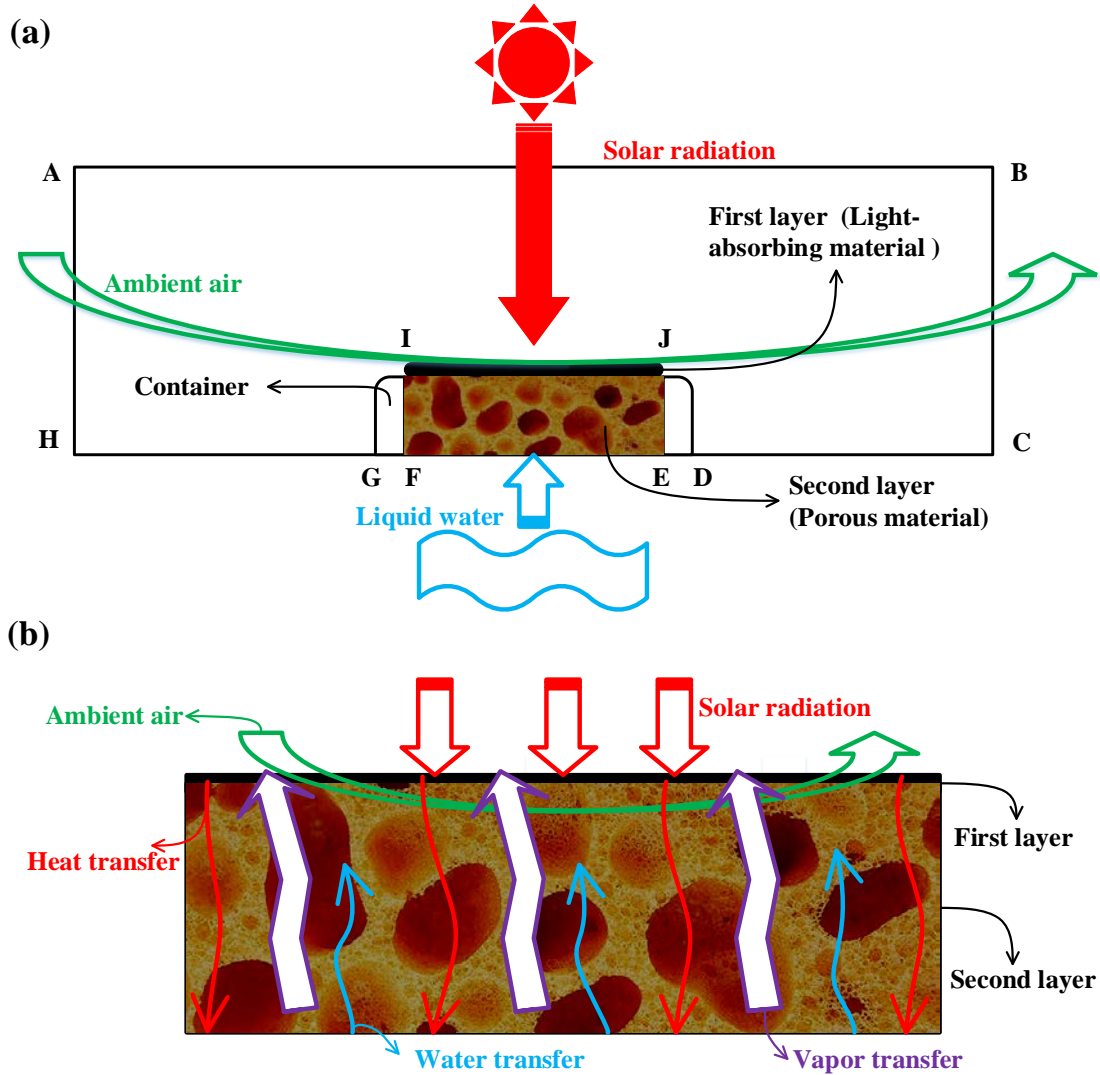

*Fig. 1 Schematic bi-layer system for solar steam generation: (a) system under solar radiation; (b) the detail of the heat and mass transfer in the second layer.*

The water evaporation in a bi-layer system is schematically shown in Fig. 1(a). In such a system, the solar radiation causes a high temperature in the first layer, and this high temperature further heats the second layer which is composed of a porous thermal insulation material for preventing the heat further transferring to the liquid water. The high temperature of the second layer could lead to vapor generation, and the capillary effect makes sure that the liquid water will flow into the second layer to replenish the water loss caused by the evaporation, as shown in Fig. 1(b). Accompanying this process, the ambient gas flows through the upside of the first layer, and the air could diffuse into the second layer and takes away the vapor. It is clear that there are three different physical processes in this system as shown in Fig. 1(b). These physical processes are further illustrated in Fig. 2 to show the air, water, vapor and heat transfers. To analyze such a system, we make the following assumptions:

**For the first layer:**
(1) Because the thickness of the first layer have a negligible effect on the evaporation

rate, here we ignore the thickness of the first layer and regard it as an optic diffuse surface;

**For the second layer:**
(2) The structure of the second layer is isotropic;
(3) Ignoring the radiation of the second layer to the environment;
(4) The viscosity dissipation like heat transfer and work caused by pressure changing is not considered to meet the assumption of local thermal equilibrium;
(5) The vapor is in equilibrium with the liquid or in other words, the time scale for evaporation is much smaller than the smallest time scale of the transport equations;
(6) Considering that the second-layer porous materials floats on the surface of the water, we assume that the second layer absorbs water only from the bottom surface;

**For ambient air:**
(7) The ambient-air flow is assumed to be laminar.

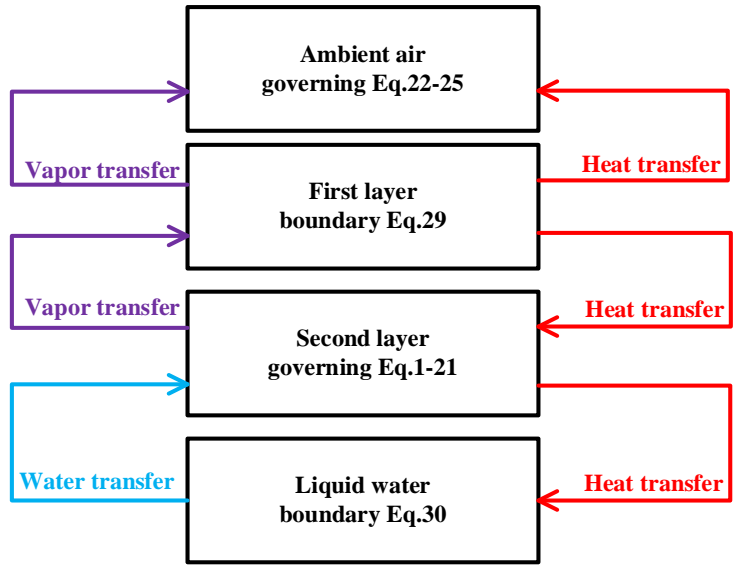

*Fig. 2 The heat and mass transfer in a bi-layer solar steam generation system.*

## 2.2 Governing equations

Under the above assumptions, the governing equations for each part of the system are given as follows.

### 2.2.1 Second layer (porous material)

The air flow in the second layer will determine the pressure gradient of gas phase. The pressure gradient accompanying with capillary effect are crucial for the water transfer. The water, air and vapor could fill all of the pores in the second layer, thus the saturation concentrations of the gas (air and vapor) and liquid phases (water) must satisfy the saturation constraint equation. In the following part, the coupled liquid and gas transfer are respectively introduced, and finally the heat transfer equation is given.

**Saturation constraint equation:**

$$S_g + S_w = 1. \tag{1}$$

where $S_g$ is the gas saturation, and the saturation of water $S_w$ can be calculated by,

$$S_w = \frac{c_w M_w}{\rho_w \varphi_p}. \tag{2}$$

where $c_w$ is the concentration of water in the second layer, which can be derived from the water transport equation.

**Water transport equation:**

$$\frac{\partial c_w}{\partial t} + \nabla \cdot (-D_{cap} \nabla c_w) + \boldsymbol{u}_w \cdot \nabla c_w = R_w, \tag{3}$$

where the capillary diffusivity $D_{cap}$ can be empirically or experimentally obtained [33], and the water velocity $\boldsymbol{u}_w$ is calculated by Darcy's Law defined in terms of the gas phase pressure gradient $\nabla P_g$,

$$\boldsymbol{u}_w = -\frac{\kappa \kappa_{rw}}{S_w \varphi_p \mu_w} \nabla P_g \boldsymbol{I}, \tag{4}$$

where the liquid relative permeability $\kappa_{rw}$ is often empirically or experimentally obtained and the value strongly depends on the properties of both porous material and liquid which has already been described in detail in Ref. [33].

Because mass is balanced between the liquid water and steam in the second layer, the reaction rate of liquid water $R_w$ in Eq.(3) is the opposite of the rate of steam generation in the second layer,

$$R_w = -K \cdot (a_w c_{sat} - c_{wg}), \tag{5}$$

where the water activity $a_w$ describes the amount of water that evaporates into air which depends on both the water content and the temperature of the surrounding air [33], and the saturation concentration $c_{sat}$ is determined by both the saturation pressure and the temperature of the second layer,

$$c_{sat} = \frac{P_{sat}}{R T_p}, \tag{6}$$

where the saturated vapor pressure $P_{sat}$ is temperature dependent which can be described by Tentens formula,

$$P_{sat} = 610.7 \times 10^{\frac{7.5 \times (T_p - 273.15)}{T - 35.85}}. \tag{7}$$

**Vapor transport equation:**

$$\frac{\partial c_{wg}}{\partial t} + \nabla \cdot (-D_{eff} \nabla c_{wg}) + \boldsymbol{u}_{wg} \cdot \nabla c_{wg} = R_{wg}, \tag{8}$$

where the effective diffusion coefficient of vapor $D_{eff}$ can be described by the Millington and Quirk equation,

$$D_{eff} = D_{va} S_g^{\frac{10}{3}} \varphi_p^{\frac{4}{3}}. \tag{9}$$

In Eq. (8), the velocity field of vapor $\boldsymbol{u}_{wg}$ is composed of the velocity fields derived from the convection diffusion and vapor diffusion,

$$\boldsymbol{u}_{wg} = \frac{\boldsymbol{u}_g}{S_g \varphi_p} - \frac{M_v D_{eff}}{M_g \rho_g} \nabla \rho_g. \tag{10}$$

In Eq. (8), the rate of steam generation $R_{wg}$ in the second layer is determined by the saturation concentration $c_{sat}$ and water activity $a_w$,

$$R_{wg} = K \cdot (a_w c_{sat} - c_{wg}). \tag{11}$$

**Gas (vapor + air) continuity equation:**

$$\nabla \cdot (\rho_g \boldsymbol{u}_g) = 0. \tag{12}$$

**Gas momentum conservation equations:**

$$\frac{\rho_g}{\varphi_p^2} \boldsymbol{u}_g \cdot \nabla \boldsymbol{u}_g = -\nabla P_g I + \nabla \cdot \left[ \frac{1}{\varphi_p} \left\{ \mu_g (\nabla \boldsymbol{u}_g + (\nabla \boldsymbol{u}_g)^T) - \frac{2}{3} \mu_g (\nabla \cdot \boldsymbol{u}_g) I \right\} \right] - \\ (\kappa \kappa_{rg})^{-1} \mu_g \boldsymbol{u}_g. \tag{13}$$

In Eqs. (12) and (13), the thermal physical properties of gas is obtained by mixing the property of vapor and dry air.

**Energy conservation equation:**

$$(\rho C_p)_e \frac{\partial T}{\partial t} + \rho_{tot} C_{p,tot} \boldsymbol{u}_{mean} \cdot \nabla T - k_e \nabla^2 T = Q_{vap}, \tag{14}$$

where the thermal physical properties of fluids (gas and water) in the second layer are average properties,

$$\rho_{tot} = S_g \rho_g + S_w \rho_w, \tag{15}$$

$$C_{p,tot} = \frac{S_g \rho_g C_{p,g} + S_w \rho_w C_{p,w}}{\rho_{tot}}, \tag{16}$$

$$k_{tot} = S_g k_g + S_w k_w. \tag{17}$$

In Eq.(14), the effective thermal properties of the porous material are described using the following equations,

$$(\rho C_p)_e = \varphi_p \rho_{tot} C_{p,tot} + (1 - \varphi_p) \rho_p C_{p,p}, \tag{18}$$

$$k_e = \varphi_p k_{tot} + (1 - \varphi_p) k_p. \tag{19}$$

The velocity field is the average velocity of moist air, water vapor and liquid water,

$$\boldsymbol{u}_{mean} = \frac{\boldsymbol{n}_a C_{p,a} + \boldsymbol{n}_{wg} C_{p,wg} + \boldsymbol{n}_w C_{p,w}}{\rho_{tot} C_{p,tot}}, \tag{20}$$

where $\boldsymbol{n}_a$, $\boldsymbol{n}_{wg}$ and $\boldsymbol{n}_w$ is respectively the mass flux for dry air, vapor and water [34]. Due to evaporation is an endothermic process, the heat of vaporization in Eq. (14) is described by the following equation,

$$Q_{vap} = -H_{vap} M_w R_{wg}. \tag{21}$$

### 2.2.2 Ambient air

Continuity equation:

$$\nabla \cdot (\rho_v \boldsymbol{u}_v) = 0. \tag{22}$$

Momentum conservation equations:

$$\rho_v(\boldsymbol{u}_v \cdot \nabla)\boldsymbol{u}_v = \nabla \cdot \left[-P_v \boldsymbol{I} + \mu_v(\nabla \boldsymbol{u}_v + (\nabla \boldsymbol{u}_v)^T) - \frac{2}{3}\mu_v(\nabla \cdot \boldsymbol{u}_v)\boldsymbol{I}\right]. \tag{23}$$

Energy conservation equation:

$$\rho_v C_{p,v} \frac{\partial T}{\partial t} + \rho_v C_{p,v} \boldsymbol{u}_v \cdot \nabla T - k_v \nabla^2 T = 0. \tag{24}$$

Vapor transport equation:

$$\frac{\partial c_v}{\partial t} + \nabla \cdot (-D_{va} \nabla c_v) + \boldsymbol{u}_v \cdot \nabla c_v = 0. \tag{25}$$

In Eqs. (22)-(25), the thermal physical properties of gas is obtained by mixing the property of vapor and dry air.

### 2.2.3 Container

Energy conservation equation:

$$\rho_s C_{p,s} \frac{\partial T}{\partial t} - k_s \nabla^2 T = 0. \tag{26}$$

### 2.3 Boundary conditions

The boundary conditions for the air flow, heat transfer, water transport, and vapor transport are respectively given in the following parts, and summarized in Table 1.

**For air flow:** The velocity at boundaries (AB, CD, EF, GH, JE, FI, JD, GI) is set to be zero. IJ is the coupling boundary between the air domain and the porous media domain. HA in Fig. 1(a) is the inlet boundary, which is defined by the Eq. (27). BC is the outlet boundary which is defined by the Eq. (28),

$$L_{in} \nabla \cdot \left[-P_v \boldsymbol{I} + \mu_v(\nabla \boldsymbol{u}_v + (\nabla \boldsymbol{u}_v)^T) - \frac{2}{3}\mu_v(\nabla \cdot \boldsymbol{u}_v)\boldsymbol{I}\right] = -P_{in}\boldsymbol{n}, \tag{27}$$

$$\left[-P_v \boldsymbol{I} + \mu_v(\nabla \boldsymbol{u}_v + (\nabla \boldsymbol{u}_v)^T) - \frac{2}{3}\mu_v(\nabla \cdot \boldsymbol{u}_v)\boldsymbol{I}\right] = -P_{out}\boldsymbol{n}. \tag{28}$$

**For heat transfer:** The boundaries (AB, CD, DE, EF, FG, GH) are defined as thermal insulation which means that there is no heat flux across the boundary. Considering that BC is the air outflow boundary, only will the heat transfer via air convection happen there. The HA is defined to be a fixed temperature boundary where the temperature is fixed as that of the inflow air. JE, FI, JD and GI are all coupled boundaries through heat flux. IJ is set as a diffuse boundary, and a radiative heat source which is defined by the Eq. (29) is added on it,

$$-\boldsymbol{n} \cdot \boldsymbol{q}_e = \varepsilon(G - e_b). \tag{29}$$

where the irradiation $G$ is calculated by $G = G_{sun} + G_{amb}$, $G_{sun} = q_e \cdot FEP(T_{sun})$, and $G_{amb} = e_b(T_{amb}) \cdot FEP(T_{amb})$. $e_b(T_{amb})$ is the blackbody hemispherical total emissive power at ambient temperature, $FEP(T)$ is the fractional blackbody emissive power at specific temperature.

**For water transport:** The pores at the EF boundary are full filled with water, and therefore the condition at the EF boundary could be defined by the Eq. (30),

$$c_w = S_{w0} \varphi_p \rho_w M_w^{-1}. \tag{30}$$

**For vapor transfer**: The inlet vapor concentration at HA is defined as $c_{v0}$. BC is outlet boundary. IJ is the coupling boundary between the air domain and the porous media domain. There is no vapor flux at the boundaries AB, BC, CD, EF, GH, JE, FI, JD, and GI.

**Table 1** Boundary conditions.

| Boundaries | Air flow | Heat transfer | Water transport | Vapor transport |
|---|---|---|---|---|
| AB | $\boldsymbol{u} = 0$ | $-\boldsymbol{n} \cdot \boldsymbol{q} = 0$ | / | $-\boldsymbol{n}(-D_{va}\nabla c_v + \boldsymbol{u}_v c_v) = 0$ |
| BC | Eq.(28) | $-\boldsymbol{n} \cdot \boldsymbol{q} = 0$ | / | $-\boldsymbol{n} \cdot D_{va}\nabla c_v = 0$ |
| CD | $\boldsymbol{u} = 0$ | $-\boldsymbol{n} \cdot \boldsymbol{q} = 0$ | / | $-\boldsymbol{n}(-D_{va}\nabla c_v + \boldsymbol{u}_v c_v) = 0$ |
| DE | / | $-\boldsymbol{n} \cdot \boldsymbol{q} = 0$ | / | / |
| EF | $\boldsymbol{u} = 0$ | $-\boldsymbol{n} \cdot \boldsymbol{q} = 0$ | Eq.(30) | $-\boldsymbol{n}(-D_{eff}\nabla c_{wg} + \boldsymbol{u}_{wg} c_{wg}) = 0$ |
| FG | / | $-\boldsymbol{n} \cdot \boldsymbol{q} = 0$ | / | / |
| GH | $\boldsymbol{u} = 0$ | $-\boldsymbol{n} \cdot \boldsymbol{q} = 0$ | / | $-\boldsymbol{n}(-D_{va}\nabla c_v + \boldsymbol{u}_v c_v) = 0$ |
| HA | Eq.(27) | $T = T_0$ | / | $c_v = c_{v0}$ |
| IJ | Coupling boundary | Eq.(29) | $-\boldsymbol{n}(-D_{cap}\nabla c_w + \boldsymbol{u}_w c_w) = 0$ | Coupling boundary |
| JE | $\boldsymbol{u} = 0$ | Coupling boundary | $-\boldsymbol{n}(-D_{cap}\nabla c_w + \boldsymbol{u}_w c_w) = 0$ | $-\boldsymbol{n}(-D_{eff}\nabla c_{wg} + \boldsymbol{u}_{wg} c_{wg}) = 0$ |
| FI | $\boldsymbol{u} = 0$ | Coupling boundary | $-\boldsymbol{n}(-D_{cap}\nabla c_w + \boldsymbol{u}_w c_w) = 0$ | $-\boldsymbol{n}(-D_{eff}\nabla c_{wg} + \boldsymbol{u}_{wg} c_{wg}) = 0$ |
| JD | $\boldsymbol{u} = 0$ | Coupling boundary | / | $-\boldsymbol{n}(-D_{va}\nabla c_v + \boldsymbol{u}_v c_v) = 0$ |
| GI | $\boldsymbol{u} = 0$ | Coupling boundary | / | $-\boldsymbol{n}(-D_{va}\nabla c_v + \boldsymbol{u}_v c_v) = 0$ |

## 3. Numerical simulation

The COMSOL Multiphysics software is used to solve the governing equations, in which the User-Defined Variable (UDV) of material properties and the User-Defined Function (UDF) of source terms are self-developed according to the mathematical models in part 2. The following size for the computational domain is applied: the lengths of AB, HA, IJ, JE, IF, and FG is respectively 90, 50, 30, 5, 5, and 1 mm. The thermal physical properties and process parameters used in the simulation are summarized in Table 2. To get a good convergence of the time dependent behavior of the system, the stationary flow equations is solved firstly by neglecting the evaporation mass source in the fluid flow computation, and then the solution is used for the time dependent study step.

**Table 2** Thermal physical properties and other parameters used in the simulations [35-37].

| Nomenclature | Value |
| --- | --- |
| Initial ambient pressure, $P_0$ | $1.01325\times10^5$ Pa |
| Initial ambient temperature, $T_0$ | 293.15 K |
| Initial ambient velocity, $\boldsymbol{u}_0$ | 0.1 m·s$^{-1}$ |
| Initial ambient vapor concentration, $c_{v0}$ | 0.2 mol·m$^{-3}$ |
| Radiation heat flux, $q_e$ | 1000 W·m$^{-2}$ |
| Molecular weight of dry air, $M_a$ | 0.028 kg·mol$^{-1}$ |
| Dry air viscosity, $\mu_a$ | $1.81\times10^{-5}$ kg·m$^{-1}$·s$^{-1}$ |
| Dry air thermal conductivity, $k_a$ | 0.025 W·m$^{-1}$·K$^{-1}$ |
| Dry air heat capacity, $C_{p,a}$ | $1.006\times10^{-5}$ J·kg$^{-1}$·K$^{-1}$ |
| Dry air density, $\rho_a$ | 1.205 kg·m$^{-3}$ |
| Molecular weight of water, $M_w$ | 0.018 kg·mol$^{-1}$ |
| Water viscosity, $\mu_w$ | $1.002\times10^{-3}$ kg·m$^{-1}$·s$^{-1}$ |
| Water thermal conductivity, $k_w$ | 0.59 W·m$^{-1}$·K$^{-1}$ |
| Water heat capacity, $C_{p,w}$ | $4.182\times10^3$ J·kg$^{-1}$·K$^{-1}$ |
| Water density, $\rho_w$ | 998.2 kg·m$^{-3}$ |
| Vapor viscosity, $\mu_{wg}$ | $1.8\times10^{-5}$ kg·m$^{-1}$·s$^{-1}$ |
| Vapor thermal conductivity, $k_{wg}$ | 0.026 W·m$^{-1}$·K$^{-1}$ |
| Vapor heat capacity, $C_{p,wg}$ | $2.062\times10^3$ J·kg$^{-1}$·K$^{-1}$ |
| Thermal conductivity of solid container, $k_s$ | 1.4 W·m$^{-1}$·K$^{-1}$ |
| Heat capacity of solid container, $C_{p,s}$ | 730 J·kg$^{-1}$·K$^{-1}$ |
| Density of solid container, $\rho_s$ | 2210 kg·m$^{-3}$ |
| Porosity of second layer, $\varphi_p$ | 0.8 |
| Thermal conductivity of porous material, $k_p$ | 0.1 W·m$^{-1}$·K$^{-1}$ |
| Heat capacity of porous material, $C_{p,p}$ | 1650 J·kg$^{-1}$·K$^{-1}$ |
| Density of porous material, $\rho_p$ | 800 kg·m$^{-3}$ |
| Overall permeability of the porous material, $\kappa$ | $1\times10^{-14}$ m$^2$ |
| Evaporation coefficient, K | 20000 l·s$^{-1}$ |
| Ideal gas constant, R | 8.314 J·mol$^{-1}$·K$^{-1}$ |
| Surface emissivity, $\varepsilon$ | 0.8 |
| Air-vapor diffusivity, $D_{va}$ | $2.6\times10^{-5}$ m$^2$·s$^{-1}$ |
| Latent heat of evaporation, $H_{vap}$ | $2.454\times10^6$ J·kg$^{-1}$ |
| Initial water saturation of second layer, $S_{iw}$ | 1 |
| Simulation time, t | 3600s |

Applied the customized mesh to boundary layers, to confirm that the simulation results will not depend on the mesh, the evaporation rate for different number of mesh grids are shown in Fig. 3 (a). It shows that when the number of grids reaches 30600, the increase of the number of grids will have a negligible effect on the solution results. In this work, the mesh with 30600 grids is applied.

The evaporation rate depends on the material properties and the evaporation process. Under the assumption that the time scale of the evaporation process is much smaller than the smallest time scale of the mass or heat transfer process, the solution of the transport equations will not depend on the value of the evaporation coefficient, and thus the evaporation rate in this work should also not depend on the value of the evaporation coefficient. It means that the value of the evaporation coefficient much be large enough to ensure that the evaporation happens immediately once the water absorbs heat. It should be also noted that the value of the evaporation coefficient should not be set to too large when a large value could impede the convergence of solutions. As shown in Fig. 3 (b), the evaporation rate converges to a constant value when the evaporation coefficient becomes larger than 20000 $l \cdot s^{-1}$. Thus, in this work the evaporation coefficient is set to 20000 $l \cdot s^{-1}$. Showing the velocity field, temperature field, vapor concentration field, and relative humidity field respectively in Fig. 4(a)-(d), it demonstrates that the numerical results for the velocity, temperature, vapor concentration and relative humidity fields are reasonable.

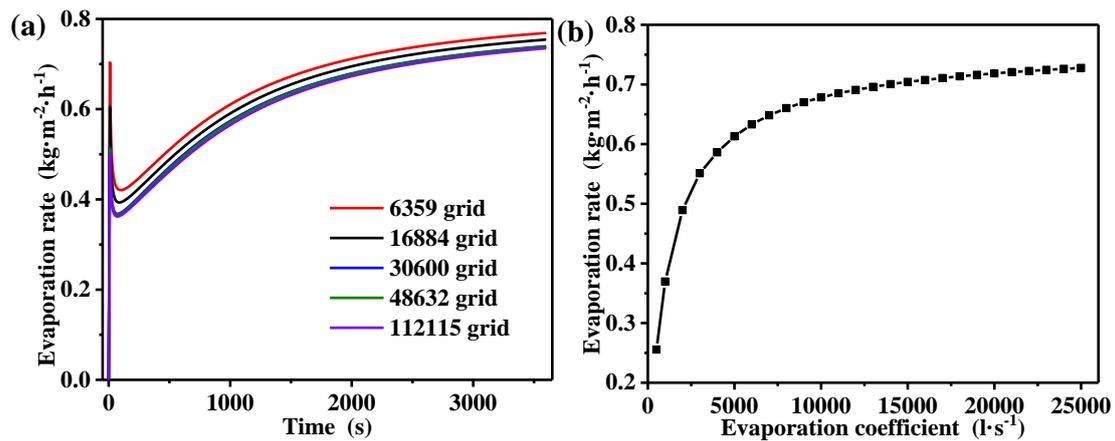

*Fig. 3 Simulation verification: (a) result independence of grids, (b) evaporation coefficient.*

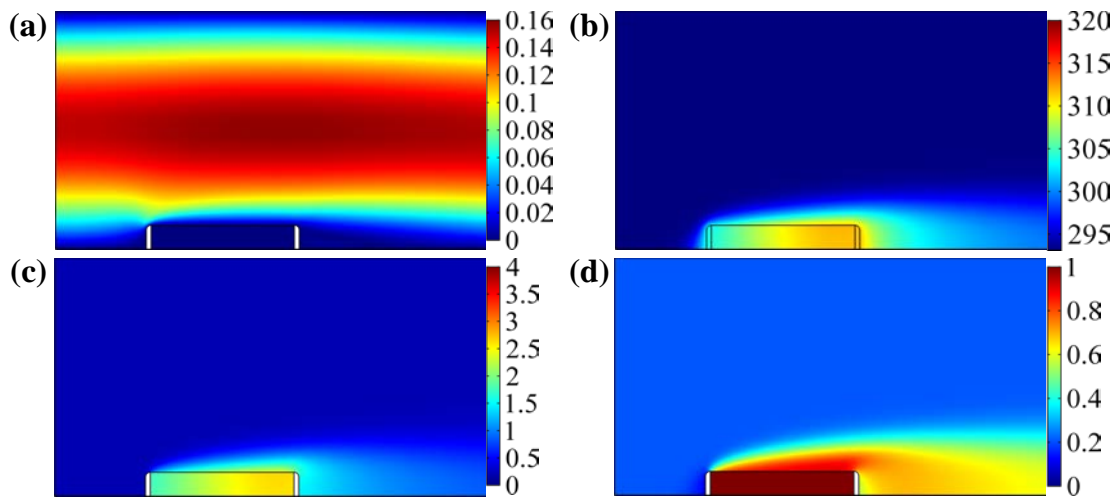

*Fig. 4 Cloud maps of simulation results: (a) velocity field, m·s$^{-1}$; (b) temperature field, K; (c) vapor concentration field, mol·m$^{-3}$; and (d) relative humidity field.*

## 4. Experiment validation

In the experiment, a cellulose sponge is used as the second layer and 50-nm copper particles are utilized as the first layer which are daubed on the second layer with a daubing density of 31.1 g·m$^{-2}$ for absorbing light. The light absorptivity of the first layer is measured with a Spectrophotometer (Lambda 750S, PerkinElmer Inc.), which is about 96%. Copper particles are commercially obtained from Beijing Dk Nano technology Co., Ltd. The thermal conductivity of the second-layer sponge with the filling air extruded out is measured to be about 0.14 W·m$^{-1}$·K$^{-1}$ with the transient hot-wire method performed on a commercial device (TC3200, Xian XIATECH Technology Co.). The porosity of the second layer is about 0.9, which is calculated by $\varphi_p = m_w/(V \cdot \rho_w)$, [38-40] where $m_w$ is the mass of the drainage water in the measurement process by an electronic balance with an uncertainty of 0.0001g, $V$ is the volume of the sponge calculated by the sectional-area multiplied by thickness, here the sectional diameter and the thickness of second layer are measured by vernier caliper with an uncertainty of 0.02 mm. $\rho_w$ is the density of water (998.2 kg·m$^{-3}$). The porosity uncertainty is calculated to be less than 0.01, small enough to be ignored. The evaporate rate is measured more than three times and a mean value is calculated with an error less than 3%.

  The experiment system is shown in Fig. 5. A solar simulator (BOS-X-350G, Bosheng Quantum Technology Co., Ltd.) is used to simulate the sun light. In the experiment, a data acquisition unit (Model 2700, Keithley Instruments, Inc.) is used to collect the temperature of the upper surface of the second layer, and an electronic analytical balance (CP214, OHAUS Instruments (Shanghai) Co., Ltd.) is applied to measure the mass change of water during evaporation. The conditions applied in the experiment are summarized in Table 3. The temperature rise at the upper surface of the second layer and the evaporation rate of the system are respectively shown in Fig. 6(a) and (b). The results got by our simulation methods are also shown for comparison. Compared to the experimental results, the deviation of the simulation result is less than 1 %, which indicates that simulation method applied here is reasonably soundable.

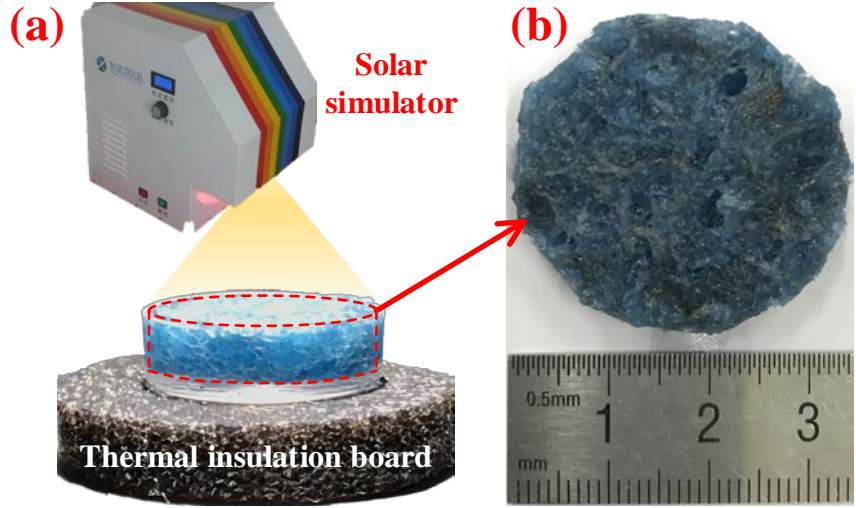

*Fig. 5 Experiment system: (a) experiment setup; (b) the bi-layer structure.*

**Table 3** Conditions in the experiment.

| Parameters | Value |
| --- | --- |
| Initial ambient velocity, $u_0$ | 0.4 m·s$^{-1}$ |
| Initial ambient vapor concentration, $c_{v0}$ | 0.431 mol·m$^{-3}$ |
| Radiation heat flux, $q_e$ | 1200 W·m$^{-2}$ |
| Surface emissivity, $\varepsilon$ | 0.96 |
| Thermal conductivity of porous material, $k_p$ | 0.14 W·m$^{-1}$·K$^{-1}$ |
| Porosity of porous material, $\varphi_p$ | 0.9 |

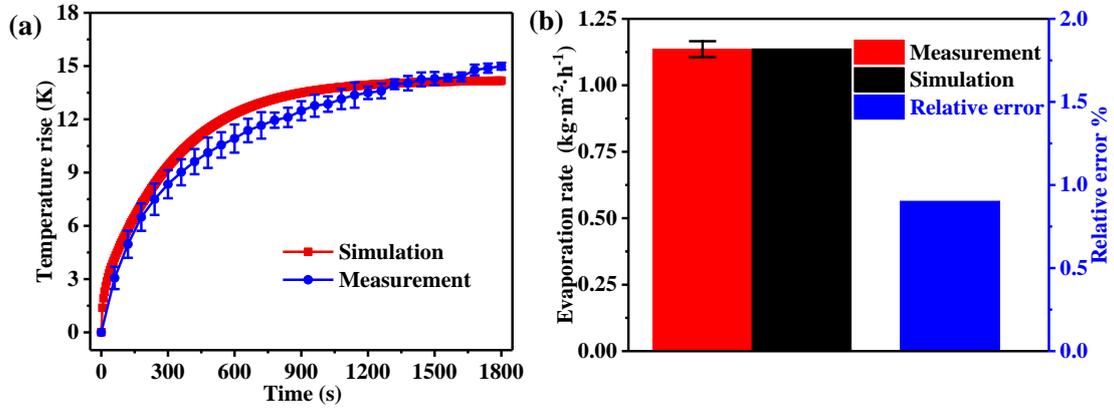

*Fig. 6 Results comparison between experimental and simulation methods: (a) the temperature rise at the upper surface of the second layer; (b) the relative error and the evaporation rate.*

## 5. Results and discussions

### 5.1 Influence of the ambient air velocity and relative humidity

It is difficult to keep the air static in a solar steam generation system, especially when the air is expected to take away the vapor. Thus, we need to consider the impact of the air velocity in the evaporation process. With the air velocity changing from 0.05 m·s$^{-1}$ to 2 m·s$^{-1}$, the simulated evaporation rate is shown in Fig. 7. It can be found that the evaporation rate increases with the rise of the air velocity, after the air velocity reaching 0.5 m·s$^{-1}$, the evaporation rate gradually tends to be a constant, as that shown in Fig. 7(a). This can be understood by that the air can take away more vapors from the second layer as the air velocity increases, and the evaporation rate tends to be a constant because the amount of vapor produced per unit time is limited by the combined effect of the porosity of the second layer and other influence factors which will be discussed in the following parts in this work. This is confirmed by that there is a lower vapor concentration in the second layer for the air velocity of 2 m·s$^{-1}$ than that for the air velocity of 0.05 m·s$^{-1}$, as that shown in Fig. 7(b) and (c).

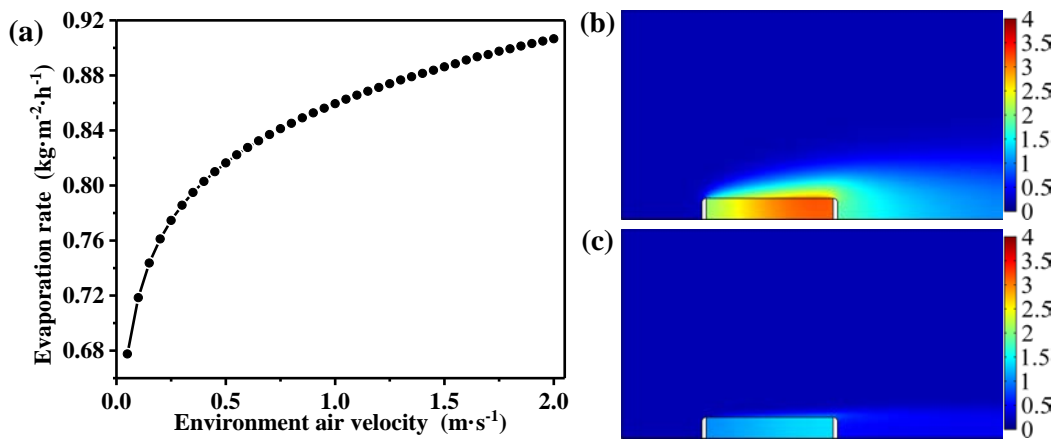

Fig. 7 Influence of the ambient air velocity: (a) evaporation rate; (b) vapor concentration field for the air velocity of 0.05 m·s$^{-1}$; (c) vapor concentration field for the air velocity of 2 m·s$^{-1}$. The unit of the vapor concentration is mol·m$^{-3}$ in (b) and (c).

The air relative humidity is also an important factor affecting the evaporation rate. The influence of the air relative humidity on the evaporation rate is shown in Fig. 8(a). It shows that the evaporation rate will decrease linearly with the increase of the relative humidity. Fig. 8 (b) and (c) respectively shows the vapor concentration field at an air relative humidity of 0 and 0.52. It illustrates that a high relative humidity in the environment could depress the rate of vapor diffusing from the second layer to the ambient air, therefore reduce the evaporation rate.

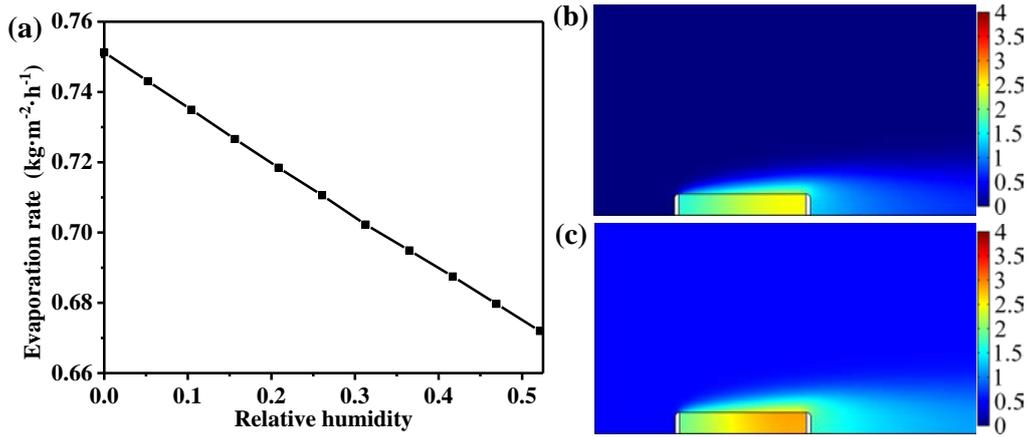

*Fig. 8 Influence of the inflow air relative humidity: (a) evaporation rate; (b) vapor concentration for the inflow air with a relative humidity of 0, (c) vapor concentration for the inflow air with a relative humidity of 0.52. The unit of the vapor concentration is mol·m$^{-3}$ in (b) and (c).*

### 5.2 Influence of the first-layer absorptivity

For the diffuse surface of the first layer, the absorptivity equals to the emissivity which can be set up for the first layer. The effect of the surface absorptivity of the first layer on the evaporation rate is studied in this part. Results are shown in Fig. 9. It shows that the evaporation rate increases almost linearly with the absorptivity increases. It means that by increasing the absorption of the first layer could greatly improve the evaporation rate. Fig. 9 (b) and (c) respectively show the vapor concentration field for the surface absorptivity of 0.4 and 0.95. It confirms that the vapor concentration in the second layer increases significantly as the surface absorptivity increases.

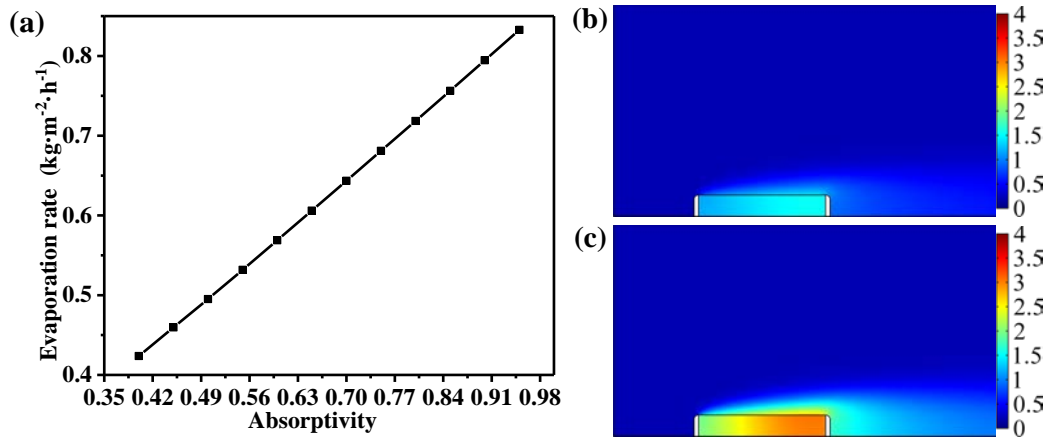

*Fig. 9 Influence of the surface absorptivity of the first layer: (a) evaporation rate; (b) the vapor concentration field for the surface absorptivity of 0.4, (c) the vapor concentration field for the surface absorptivity of 0.95. The unit of the vapor concentration is mol·m$^{-3}$ in (b) and (c).*

### 5.3 Influence of the second-layer porosity and thermal conductivity

The porosity of the second layer is a key factor affecting the evaporation rate. In this part, the effect of the porosity of the second layer on the evaporation rate is focused. Results are shown in Fig. 10(a). The evaporation rate firstly increases rapidly with the porosity increasing from 0.2 to 0.4, and then gradually decreases. There is an optimum porosity for obtaining a high evaporation rate. To understand the underlying mechanism of the existence of the optimum porosity, the vapor concentration field at different porosities, such as 0.2, 0.45, and 0.95, are respectively shown in Fig. 10 (b)-(d). When a low porosity of 0.2 is applied, there is little water could be hold in pores and thus the effective thermal conductivity of the second layer will be low. This low effective thermal conductivity of the second-layer could result in a higher temperature, thereby generating more vapor in the second-layer porous material, but the vapor is hard to diffuse to the air because of the low porosity. When a high porosity of 0.9 is applied, the high effective thermal conductivity of the second layer could lead to a fast temperature loss, and thus a decrease of the vapor production. The largest evaporation rate occurring at a porosity of about 0.45 should arrive from the balanced influence of the positive effect of the lowering effective thermal conductivity and the negative effect of the decreasing vapor diffusivity when the porosity is decreased in the second layer.

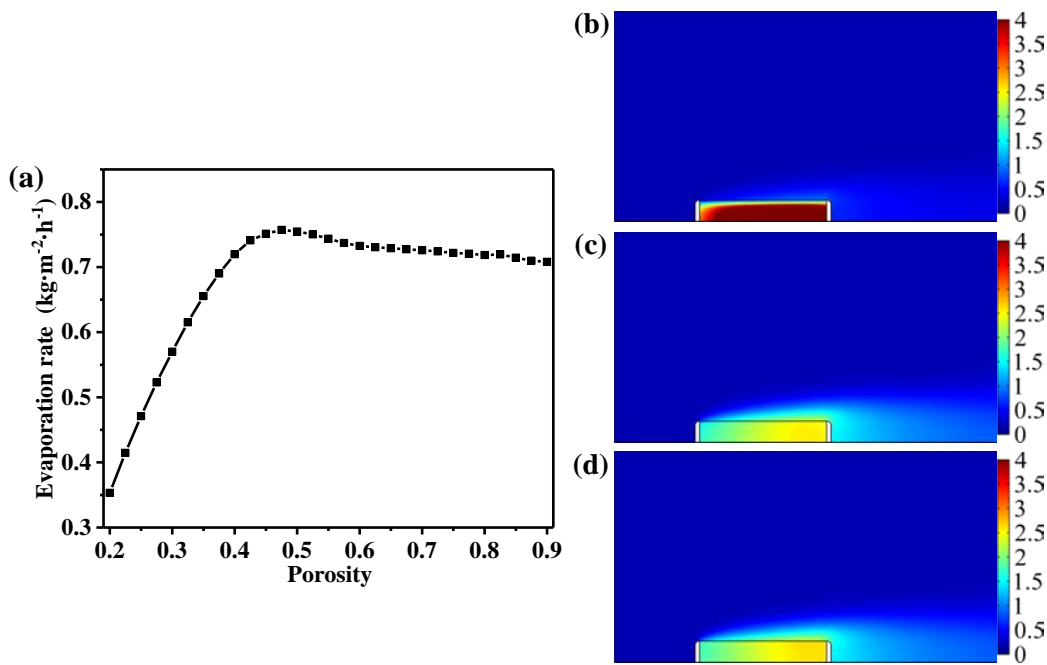

*Fig. 10 Influence of the second-layer porosity: (a) evaporation rate; (b) the vapor concentration field at the porosity of 0.2; (c) the vapor concentration field at the porosity of 0.45, (d) the vapor concentration field at the porosity of 0.9. The unit of the vapor concentration is mol·m$^{-3}$ in (b), (c) and (d).*

The effect of the thermal conductivity of the second-layer porous material on the evaporation rate is taken into account in this part. Fig. 11(a) depicts that the effect of the thermal conductivity of the porous material on the evaporation rate is negligible, whether a high porosity of 0.8 or a low porosity of 0.45 for porous material is applied. Fig. 11(b) and (c) respectively show the vapor concentration field for a thermal conductivity of 0 or 10 W·m⁻¹·K⁻¹ of the porous material with the porosity of 0.8. It is obvious that the vapor concentration field is almost same for the thermal conductivity of 0 and 10 W·m⁻¹·K⁻¹, which confirms that the thermal conductivity of the porous material has a negligible effect on the evaporation rate. The negligible effect of the thermal conductivity of the porous material is easy to be understood by that: the effective thermal conductivity of the second layer, which could determine the heat transfer from the first layer to liquid water, mainly depends on the porosity not the thermal conductivity of the porous material, because a larger porosity could lead to more water in the second layer, and more water will result in a larger heat transfer ability and thus a small temperature rise.

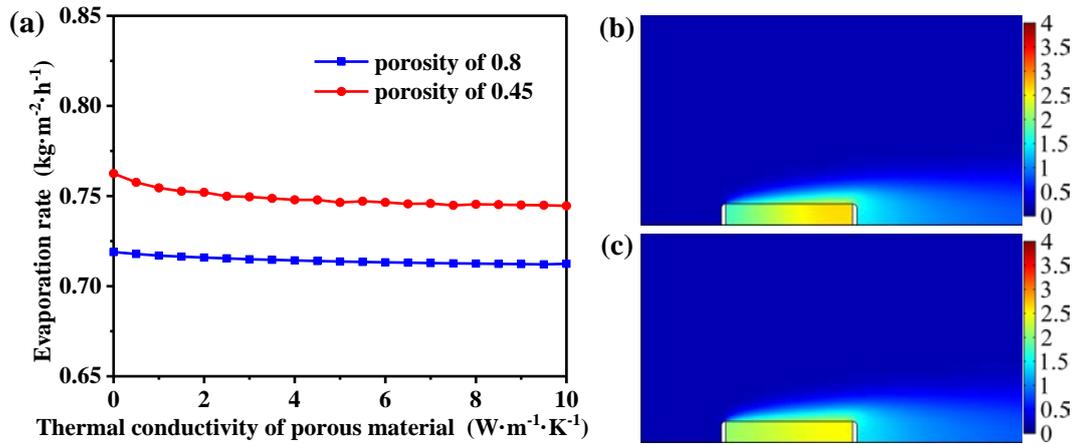

Fig. 11 Influence of the thermal conductivity of the porous material: (a) evaporation rate, (b) vapor concentration field for the thermal conductivity of 0 W·m⁻¹·K⁻¹, (c) vapor concentration field for the thermal conductivity of 10 W·m⁻¹·K⁻¹. The unit of the vapor concentration is mol·m⁻³ in (b) and (c).

## 6. Conclusions

A mathematical model has been developed to analyze the evaporation rate of the bi-layer system. The deviation of the model result from the experimental one is less than 1 %. The effects of the ambient air velocity, the air relative humidity, the absorptivity of the first-layer material, the porosity and the thermal conductivity of the second-layer material are respectively considered. Results turn out that: (1) the evaporation rate increases with the rise of the ambient air velocity until the air velocity reaching 0.5 m·s⁻¹, then the evaporation rate gradually approaches a constant value; (2) The evaporation rate decreases linearly with the increase of the air relative humidity, because a high relative humidity could depress the rate of vapor diffusing from the second layer to air; (3) The evaporation rate increases almost linearly with the increase of the absorptivity of the first layer; (4) There is an optimum porosity for the second layer to enhance the evaporation rate, and the optimum porosity of about 0.45 is obtained in this work; (5) The effect of the thermal conductivity of the second-layer porous material on the evaporation rate is negligible, because the effective thermal conductivity of the second layer is determined by the porosity.

## Acknowledgment

This work has been supported by the Fundamental Research Funds for the Central Universities (2018XKQYMS17).

## References

[1] R. Agrawal, N.R. Singh, F.H. Ribeiro, W.N. Delgass, Sustainable fuel for the transportation sector, Proceedings of the National Academy of Sciences of the United States of America, 104(12) (2007) 4828-4833.
[2] E. Cartlidge, SAVING FOR A RAINY DAY, Science, 334(6058) (2011) 922-924.
[3] M. Elimelech, W.A. Phillip, The Future of Seawater Desalination: Energy, Technology, and the Environment, Science, 333(6043) (2011) 712-717.
[4] M.K. Gupta, S.C. Kaushik, Exergy analysis and investigation for various feed water heaters of direct steam generation solar-thermal power plant, Renewable Energy, 35(6) (2010) 1228-1235.
[5] M.A. Shannon, P.W. Bohn, M. Elimelech, J.G. Georgiadis, B.J. Marinas, A.M. Mayes, Science and technology for water purification in the coming decades, Nature, 452(7185) (2008) 301-310.
[6] E. Zarza, L. Valenzuela, J. Leon, K. Hennecke, M. Eck, H.D. Weyers, M. Eickhoff, Direct steam generation in parabolic troughs: Final results and conclusions of the DISS project, Energy, 29(5-6) (2004) 635-644.
[7] F. Wang, L. Ma, Z. Cheng, J. Tan, X. Huang, L. Liu, Radiative heat transfer in solar thermochemical particle reactor: A comprehensive review, Renewable & Sustainable Energy Reviews, 73 (2017) 935-949.
[8] G. Wei, G. Wang, C. Xu, X. Ju, L. Xing, X. Du, Y. Yang, Selection principles and thermophysical properties of high temperature phase change materials for thermal energy storage: A review, Renewable & Sustainable Energy Reviews, 81 (2018) 1771-1786.
[9] H. Ding, G. Peng, S. Mo, D. Ma, S.W. Sharshir, N. Yang, Ultra-fast vapor generation by a graphene nano-ratchet: a theoretical and simulation study, Nanoscale, 9(48) (2017).
[10] H. Ghasemi, G. Ni, A.M. Marconnet, J. Loomis, S. Yerci, N. Miljkovic, G. Chen, Solar steam generation by heat localization, Nature Communications, 5 (2014).
[11] C. Dong, W. Ma, X. Zhang, Modulation of the time-domain reflectance signal induced by thermal transport and acoustic waves in multilayered structure, International Journal of Heat and Mass Transfer, 114 (2017) 915-922.
[12] W. Ma, T. Miao, X. Zhang, K. Takahashi, T. Ikuta, B. Zhang, Z. Ge, A T-type method for characterization of the thermoelectric performance of an individual free-standing single crystal $Bi_2S_3$ nanowire, Nanoscale, 8(5) (2016) 2704-2710.
[13] Y. Liu, S. Yu, R. Feng, A. Bernard, Y. Liu, Y. Zhang, H. Duan, W. Shang, P. Tao, C. Song, T. Deng, A Bioinspired, Reusable, Paper-Based System for High-Performance Large-Scale Evaporation, Advanced Materials, 27(17) (2015) 2768-2774.

[14] X. Zhao, C. Huang, Q. Liu, I.I. Smalyukh, R. Yang, Thermal conductivity model for nanofiber networks, Journal of Applied Physics, 123(8) (2018).
[15] Y. Ito, Y. Tanabe, J. Han, T. Fujita, K. Tanigaki, M. Chen, Multifunctional Porous Graphene for High-Efficiency Steam Generation by Heat Localization, Advanced Materials, 27(29) (2015) 4302-4307.
[16] S.W. Sharshir, G. Peng, L. Wu, F.A. Essa, A.E. Kabeel, N. Yang, The effects of flake graphite nanoparticles, phase change material, and film cooling on the solar still performance, Applied Energy, 191 (2017) 358-366.
[17] Z.-Z. Lin, C.-L. Huang, Z. Huang, W.-K. Zhen, Surface/interface influence on specific heat capacity of solid, shell and core-shell nanoparticles, Applied Thermal Engineering, 127 (2017) 884-888.
[18] Y. Liu, J. Chen, D. Guo, M. Cao, L. Jiang, Floatable, Self-Cleaning, and Carbon-Black-Based Superhydrophobic Gauze for the Solar Evaporation Enhancement at the Air-Water Interface, Acs Applied Materials & Interfaces, 7(24) (2015) 13645-13652.
[19] W. Ma, T. Miao, X. Zhang, L. Yang, A. Cai, Z. Yong, Q. Li, Thermal performance of vertically-aligned multi-walled carbon nanotube array grown on platinum film, Carbon, 77 (2014) 266-274.
[20] H. Li, Y. He, Y. Hu, X. Wang, Commercially Available Activated Carbon Fiber Felt Enables Efficient Solar Steam Generation, ACS applied materials & interfaces, 10(11) (2018) 9362-9368.
[21] X. Wang, Y. He, X. Liu, J. Zhu, Enhanced direct steam generation via a bio-inspired solar heating method using carbon nanotube films, Powder Technology, 321 (2017) 276-285.
[22] F. Wang, Y. Shuai, H. Tan, C. Yu, Thermal performance analysis of porous media receiver with concentrated solar irradiation, International Journal of Heat and Mass Transfer, 62 (2013) 247-254.
[23] F. Wang, Y. Shuai, H. Tan, X. Zhang, Q. Mao, Heat transfer analyses of porous media receiver with multi-dish collector by coupling MCRT and FVM method, Solar Energy, 93 (2013) 158-168.
[24] G. Wei, P. Huang, C. Xu, L. Chen, X. Ju, X. Du, Experimental study on the radiative properties of open-cell porous ceramics, Solar Energy, 149 (2017) 13-19.
[25] F. Wang, J. Tan, Y. Shuai, H. Tan, S. Chu, Thermal performance analyses of porous media solar receiver with different irradiative transfer models, International Journal of Heat and Mass Transfer, 78 (2014) 7-16.
[26] C.-L. Huang, Z.-Z. Lin, Y.-H. Feng, X.-X. Zhang, G. Wang, Thermal conductivity prediction of 2-dimensional square-pore metallic nanoporous materials with kinetic method approach, International Journal of Thermal Sciences, 112 (2017) 263-269.
[27] K.-K. Liu, Q. Jiang, S. Tadepallifit, R. Raliya, P. Biswas, R.R. Naik, S. Singamaneni, Wood Graphene Oxide Composite for Highly Efficient Solar Steam Generation and Desalination, Acs Applied Materials & Interfaces, 9(8) (2017) 7675-7681.
[28] H. Liu, C. Chen, G. Chen, Y. Kuang, X. Zhao, J. Song, C. Jia, X. Xu, E. Hitz, H. Xie, S. Wang, F. Jiang, T. Li, Y. Li, A. Gong, R. Yang, S. Das, L. Hu, High-Performance Solar Steam Device with Layered Channels: Artificial Tree with a Reversed Design,

Advanced Energy Materials, 8(8) (2018).
[29] G. Xue, K. Liu, Q. Chen, P. Yang, J. Li, T. Ding, J. Duan, B. Qi, J. Zhou, Robust and Low-Cost Flame-Treated Wood for High-Performance Solar Steam Generation, Acs Applied Materials & Interfaces, 9(17) (2017) 15052-15057.
[30] N. Xu, X. Hu, W. Xu, X. Li, L. Zhou, S. Zhu, J. Zhu, Mushrooms as Efficient Solar Steam-Generation Devices, Advanced Materials, 29(28) (2017).
[31] F. Jiang, H. Liu, Y. Li, Y. Kuang, X. Xu, C. Chen, H. Huang, C. Jia, X. Zhao, E. Hitz, Y. Zhou, R. Yang, L. Cui, L. Hu, Lightweight, Mesoporous, and Highly Absorptive All-Nanofiber Aerogel for Efficient Solar Steam Generation, Acs Applied Materials & Interfaces, 10(1) (2018) 1104-1112.
[32] Q. Jiang, S. Singamaneni, Water from Wood: Pouring through Pores, Joule, 1(3) (2017) 429-430.
[33] A.K. Datta, Porous media approaches to studying simultaneous heat and mass transfer in food processes. II: Property data and representative results, Journal of Food Engineering, 80(1) (2007) 96-110.
[34] A.K. Datta, Porous media approaches to studying simultaneous heat and mass transfer in food processes. I: Problem formulations, Journal of Food Engineering, 80(1) (2007) 80-95.
[35] T.L. Bergman, F.P. Incropera, A.S. Lavine, D.P. DeWitt, Introduction to heat transfer, John Wiley & Sons, 2011.
[36] E.R.G. Eckert, R.M. Drake Jr, Analysis of heat and mass transfer, (1987).
[37] E.L. Cussler, Diffusion: Mass Transfer in Fluid Systems, Cambridge University Press, 2009.
[38] D. Wu, C. Huang, J. Zhong, Z. Lin, Influence factors of the inter-nanowire thermal contact resistance in the stacked nanowires, Physica B-Condensed Matter, 537 (2018) 150-154.
[39] Z.-Z. Lin, C.-L. Huang, D.-C. Luo, Y.-H. Feng, X.-X. Zhang, G. Wang, Thermal performance of metallic nanoparticles in air, Applied Thermal Engineering, 105 (2016) 686-690.
[40] Z.-Z. Lin, C.-L. Huang, W.-K. Zhen, Z. Huang, Enhanced thermal conductivity of metallic nanoparticle packed bed by sintering treatment, Applied Thermal Engineering, 119 (2017) 425-429.